\documentstyle[multicol,epsf,aps]{revtex}
\def\ruleleft{\vspace{-2.5\baselineskip}\begin{multicols}{2}\ \linebreak\vspace{-\baselineskip}\hrulefill\raisebox
{0.84mm}{$\!\rfloor$}\[\]\end{multicols}\vspace{-1.5\baselineskip}}
\def\ruleright{\vspace{-1.5\baselineskip}\begin{multicols}{2}\ \linebreak\raisebox
{-2.45mm}{$\lceil\!$}\hrulefill\end{multicols}\vspace{-\baselineskip}}
\newlength{\normalarraycolsep}
\newlength{\normaltabcolsep}
\newcommand{\bm}[1]{\mbox{\boldmath $#1$}}
\newcommand{\bms}[1]{\mbox{\scriptsize \boldmath $#1$}}
\newcommand{\rb}{\bm{r}}
\newcommand{\Eb}{\bm{E}}
\newcommand{\fracd}[2]{\frac{\displaystyle#1}{\displaystyle#2}}
\newcommand{\be}{\begin{equation}}
\newcommand{\ee}{\ \ \end{equation}}
\newcommand{\bea}{\setlength{\arraycolsep}{0.4\normalarraycolsep}
                  \begin{eqnarray}} 
\newcommand{\eea}{\ \ \end{eqnarray}\setlength{\arraycolsep}
                  {\normalarraycolsep}}
\newcommand{\bean}{\setlength{\arraycolsep}{0.4\normalarraycolsep} 
                  \begin{eqnarray*}}
\newcommand{\eean}{\ \ \ \end{eqnarray*}\setlength{\arraycolsep}
                  {\normalarraycolsep}}

\begin{document}
 \title{\bf Debye-H\"uckel-Bjerrum theory for charged colloids}
\author{{\bf M. N. Tamashiro, Yan Levin\footnote{Corresponding author.}
 and Marcia C. Barbosa}}
\address {\it Instituto de F\'{\i}sica,
Universidade Federal do Rio Grande do Sul\\
 Caixa Postal 15051, 91501-970 Porto Alegre (RS), Brazil \\
{\small mtamash@if.ufrgs.br, levin@if.ufrgs.br, barbosa@if.ufrgs.br}}
\maketitle 
\begin{abstract}
We formulate an extension of the Debye-H\"uckel-Bjerrum theory
[M. E. Fisher and Y. Levin, Phys. Rev. Lett. {\bf 71}, 3826 (1993)]
to the fluid state of a highly asymmetric charged colloid.
Allowing for the formation of clusters consisting of one polyion and 
$n$ condensed counterions, the total Helmholtz free energy of the 
colloidal suspension is constructed. The thermodynamic properties, 
such as the cluster-density distribution and the pressure, are obtained 
by the minimization of the free energy 
under the constraints of fixed number of polyions and counterions.  
In agreement with the current 
experimental and Monte Carlo results, 
no evidence of any phase transition is encountered.
\end{abstract}
\bigskip

\centerline{{\bf PACS numbers:} 82.70.Dd; 36.20.$-$r; 64.60.Cn}
\bigskip

\begin{multicols}{2}

\section{Introduction}

The technological importance of charged colloidal suspensions can not
be overemphasized. One comes across these important systems in fields as
diverse as the chemical engineering and the environmental science. Many 
water-soluble paints contain charged colloidal suspensions as a main 
ingredient. In this case the problem of great industrial importance
is to stabilize the suspension against the flocculation and 
precipitation. On the other extreme is the constantly growing environmental 
necessity of cleaning contaminated water. For this it is essential to find
the 
most effective way of precipitating the (usually) charged organic 
particles dissolved in the water. 

From the theoretical perspective the problem of strongly asymmetric 
electrolyte solutions is extremely difficult to study. The long-ranged 
nature of the Coulomb force combined with the large charge and size 
asymmetry between the polyions and the counterions and coions makes 
it impossible to use the traditional methods of liquid-state theory.
At high volume fractions the suspension will crystallize, that is, the 
polyions will become arranged in the form of a lattice. The solid 
state provides a  major simplification of reducing the many-polyion 
problem to that of 
one polyion inside a Wigner-Seitz (WS) cell \cite{Ma55}. 
Unfortunately, at low 
densities, or in the presence of a simple electrolyte, the suspension 
becomes disordered and the WS picture is no longer valid \cite{Lo93}. 
A new strategy 
must be tried. 

The Debye-H\"uckel-Bjerrum theory (DHBj) \cite{Fi93} 
was quite successful in explaining the behavior of symmetric 
electrolytes. The Bjerrum's
concept of association of oppositely charged ions into 
dipolar pairs \cite{Bj26} 
served to correct the Debye-H\"uckel (DH)
linearization of
the nonlinear Poisson-Boltzmann equation \cite{DH23}. 
Taking into account
the dipolar solvation energy made the coexistence curve produced by
the DHBj theory \cite{Fi93}
become in excellent agreement with the Monte Carlo
simulations \cite{Pa92}.  
The large surface charge of a polyion suggests that the cluster formation
should be even more important in the case of polyelectrolytes. In the 
present work we shall explore to what extent the counterion 
condensation influences the thermodynamic properties of a polyelectrolyte 
solution. The DHBj theory will be extended to treat the fluid 
state of a charged colloidal
suspension.

\section{Definition of the model}

We shall work with the primitive model of polyelectrolyte 
(PMP) \cite{LBT97}. 
The system will consist
 of $N_p$ polyions inside a volume $V$. The polyions are 
modeled as hard spheres of radius $a$ with a uniform surface charge
density 
\be
\sigma_0=-\frac{Zq}{4\pi a^2} \:,
\ee
where $Z$ is the polyion valence (number 
of ionized sites) and $q>0$ is the charge of a proton.
To preserve the overall charge neutrality of the system, $Z N_p$ 
point-like counterions of charge $+q$ are present. 
The solvent is modeled as a homogeneous medium of dielectric constant $D$.
Due to the strong electrostatic interaction between the polyions and the
counterions, we expect that the asymmetric  polyelectrolyte will be 
composed of clusters, with density $\rho_n$, consisting of one polyion 
and  $1\leq n \leq Z$ associated counterions, as well as bare polyions 
of density $\rho_{0}$ and free (unassociated) counterions of 
density $\rho_{f}$. The conservation of the total number of polyions and
counterions leads to two conservation equations,
\be
\rho_p=\sum_{n=0}^{Z} \rho_{n} \:,
\ee
and 
\be
Z\rho_{p}=\rho_{f}+\sum_{n=0}^{Z} 
n\rho_{n}\:,
\ee 
where $\rho_p=N_{p}/V$ is
the total density of polyions (associated or not). 

All the thermodynamic properties of the system
can be determined once the free energy is known.
The Helmholtz free energy can be split into two parts,
the electrostatic and entropic. The electrostatic terms are due 
to the inter-particles interactions and
can be attributed to the polyion-counterion, 
the polyion-polyion, and the counterion-counterion interactions. All the
electrostatic interactions will be evaluated using the DH 
theory \cite{DH23}. This is motivated by the former success
of the theory when it was 
applied to symmetric electrolytes. In principle, the linear DH 
theory is satisfactory only for low densities and high temperatures. 
However, once the concept of clusters is introduced, the validity of the 
DHBj theory is extended into the nonlinear regime \cite{Fi93}. 
The linear structure of the DHBj theory insures its internal self 
consistency, a problem 
which is intrinsic to many of the nonlinear theories of electrolyte 
solutions \cite{On33}.  

We shall assume 
that the effect of the counterion condensation is to renormalize the 
polyion charge. Thus, the effective surface charge of a $n$-cluster is
\be 
\sigma_n= -\frac{(Z-n)q}{4\pi a^2} = \sigma_0 \frac{Z-n}{Z} \: . 
\ee
All the nonlinearities related to the internal 
degrees of freedom of the clusters will be included in the entropic terms. 
In a previous work [7a]
we have considered that the 
bounded counterions condense onto the surface of the spherical polyion.
Although less realistic, this assumption has allowed us to obtain 
closed analytical expressions for the entropic contribution. 
In the present work, however, the intra-cluster interactions will 
be treated using a local-density functional theory, 
so that the correlations between the bound counterions are explicitly 
taken into account. These correlations effects can be disregarded only
when the concentration of counterions is not too high, a condition which 
may  not be fulfilled in the close vicinity of a highly charged polyion.
We now proceed to describe each one of the 
contributions to the Helmholtz free-energy density $f=-F/V$.

\section{The polyion-counterion interaction}

The {\it polyion-counterion}\/ contribution is obtained using the usual
DH theory applied to a $n$-cluster of effective surface
charge $\sigma_n$ inside the
ionic atmosphere [7a]. Consider a $n$-cluster fixed at the origin, $r=0$.
Due to the hard-sphere exclusion, no counterions will be found inside the 
region $r<a$, that is, 
\be
\rho_q (r<a) =0 \:. 
\ee 
Outside the spherical polyion, $r\ge a$, the cluster-counterion correlation
function is approximated by a Boltzmann factor, leading to the charge
density
\bea
\rho_q(r\ge a)&=&q\rho_f \exp\left[-\beta q \psi_n(r)\right] 
- \sum_{n=0}^Z (Z-n)q\rho_n \nonumber\\
&&  + \sigma_n \, \delta (r-a) \:,  \label{eqn:rho>}
\eea
where $\beta^{-1}=k_{\rm B} T$ and $\psi_n(r)$ is the electrostatic potential
at a distance $r$ from the center of the polyion. 
Notice that only the free counterions are assumed to get polarized; the 
bare polyions and clusters are too massive to be affected 
by the electrostatic fluctuations and contribute only to the neutralizing
background. Substituting the charge density into the Poisson equation,
\be 
\nabla^2 \psi_n (r) = -\frac{4\pi}{D} \rho_q(r) \: ,
\ee
one obtains the nonlinear Poisson-Boltzmann equation. 
After the linearization of the exponential factor in (\ref{eqn:rho>}), 
the electrostatic potential $\psi_n(r)$
satisfies the Laplace (for $r<a$) and the Helmholtz (for $r\ge a$) 
equations,
\be
\nabla^2 \psi_n(r) = 
\left\{
\begin{array}{cl}
0\: , & \  \mbox{ for } r<a\:, \\
\kappa^2 \psi_n(r) -\fracd{4\pi}{D}\sigma_n \, \delta (r-a) \:, & 
\  \mbox{ for } r\ge a\:, 
\end{array} 
\right.
\ee
where $\kappa=\sqrt{4\pi\lambda_{\rm B} \rho_f}$ is the inverse Debye screening
length, and $\lambda_{\rm B}=\beta q^2/D$ is the Bjerrum length. In principle
the linearization is valid only in the limit $\beta q \psi_n \ll 1$,
however, since the formation of clusters is taken into account,
the validity of the theory is extended into the nonlinear 
regime \cite{Fi93}.

The second-order differential equation for $\psi_n(r)$ can be integrated,
supplemented by the boundary conditions of vanishing of the electrostatic 
potential at infinity, the continuity of $\psi_n(r)$ at $r=a$, 
and the discontinuity in the radial component of the electric field related 
to the presence of the surface charge $\sigma_n$ at $r=a$. 
Under these conditions, we obtain
\be
\psi_n(r) = 
\left\{
\begin{array}{cl}
-\fracd{(Z-n)q}{Da(1+\kappa a)}\:, & \ \mbox{ for } r<a\:, \\
-\fracd{(Z-n)qe^{\kappa(a-r)}}{Dr(1+\kappa a)}\:, & \ \mbox{ for } 
r\ge a\:.
\end{array} 
\right.
\ee
Using the charge density in the linearized form,
\be
\rho_q(r) = 
\left\{
\begin{array}{cl}
0\:, & \ \mbox{ for } r<a\:, \\
-\fracd{\kappa^2 D}{4\pi} \psi_n(r) + \sigma_n \, \delta (r-a) \:,
& \ \mbox{ for } r\ge a\:,
\end{array} 
\right.
\ee
the electrostatic energy of a $n$-cluster is calculated to be
\bea
U_n(\kappa,q)&=& \frac12 \int d^3\rb\, \rho_q(r) \psi_n(r)
\nonumber\\
& =& 
\fracd{(Z-n)^2 q^2}{D(1+\kappa a)} 
\left[ \frac1a -\fracd{\kappa}{2(1+\kappa a)} \right]\:. 
\eea
The electrostatic free-energy density for the polyion-counterion 
interaction is obtained through the Debye charging process, in which all
the particles are simultaneously charged from 0 to their final 
charge \cite{DH23}, 
\bea
\beta {f}^{\rm PC} (\rho_f,\{ \rho_n\} )&=&
-\sum_{n=0}^Z \rho_n \int_0^1 d\lambda\,
\frac{2\beta U_n (\lambda \kappa, \lambda q)}{\lambda} \nonumber\\
&=&   
-\sum_{n=0}^Z\frac{(Z-n)^2 \lambda_{\rm B}}{2a(1+\kappa a)} \rho_n \:.
\eea
  
\section{The polyion-polyion interaction}

Due to the large asymmetry between the polyions and the counterions,
the degrees of freedom associated with the counterions can be effectively 
integrated out. 
The long-ranged interaction between two clusters will be screened by 
the cloud of free counterions,  producing an effective short-ranged 
potential of a DLVO form \cite{DLVO,So84},
\be
V^{\rm eff}_{nm} (r)
= q^2(Z-n)(Z-m)\frac{\exp(2\kappa a - \kappa r)}
{D r (1+\kappa a)^2} \:. \ee 
The {\it polyion-polyion}\/ contribution to the 
Helmholtz free energy can then be calculated in the spirit
of the usual van-der-Waals theory [3,7a],
\end{multicols}
\ruleleft
\medskip

\bea
\beta {f}^{\rm PP}(\rho_f)&=&
-\frac12\sum_{n=0}^{Z}\sum_{m=0}^{Z}
\beta\rho_n\rho_m\int d^3\rb \, V^{\rm eff}_{nm} (r) \nonumber \\
&=& -2 \pi \lambda_{\rm B}
\frac{(1+2\kappa a)}{\kappa^2 \left(1+\kappa a\right)^2} 
\sum_{n=0}^{Z}\sum_{m=0}^{Z} (Z-n)(Z-m) \rho_n\rho_m= 
-2 \pi \lambda_{\rm B}\frac{(1+2\kappa a)}{\kappa^2 \left(1+\kappa a\right)^2}
 \rho_f^2 \:.
\eea
 
\ruleright
\begin{multicols}{2}

\section{The counterion-counterion interaction}

The {\it counterion-counterion}\/ contribution, originating from the
interactions between the {\it free}\/ counterions, is calculated 
using the One Component Plasma (OCP) theory \cite{No84}. 
The electrostatic free energy is found through a Debye charging process
and a closed analytic form for $f^{\rm CC}(\rho_f)$, valid 
over a wide range of densities, is presented in [11b],
\bea
\beta f^{\rm CC}(\rho_f) &=& -\rho_f {\cal F}_{\rm corr} (\rho_f) \:, \\
{\cal F}_{\rm corr} (\rho_f) &=&
 \frac14\left[ 1+ \frac{2\pi}{3\sqrt{3}} + 
\ln\left(\frac{\omega^2+\omega+1}{3}\right)-\omega^2\right.\nonumber\\
&& \left. -
\frac{2}{\sqrt{3}}\tan^{-1}\left(\frac{2\omega+1}{\sqrt{3}}\right)\right]
\:, \label{eqn:ocp1} \\
\omega&=& \omega(\rho_f) =
\left\{ 1+ 3 \left[4\pi\lambda_{\rm B}^3 \rho_f \right]^{1/2} \right\}^{1/3} 
\:. \label{eqn:ocp2}
\eea
In the bulk this contribution is very small, and is included only for
completeness.

\section{The mixing free energy}

The mixing free energy reduces to a sum of ideal-gas terms,
\be
\beta{f}^{\rm mix} (\rho_f,\{\rho_n\}) = 
\sum_{s} \rho_{s} \left[ 1 - 
\ln \left( \rho_{s} \Lambda_s^3/ \zeta_s \right) \right] \:,
\ee
where $s\in \{f ; n=0,\ldots,Z \}$, $\Lambda_s$ are the thermal 
de Broglie wavelengths associated with free counterions, bare polyions,
and clusters; $\zeta_s$ are the internal partition functions for an 
isolated specie $s$.
Since the bare polyions and the free (unassociated) counterions do not 
have internal structure, their internal partition functions are  simply  
given by $\zeta_0=\zeta_f=1$. For a $n$-cluster the internal partition 
function is
\bea
\zeta_n&=&
 \frac{1}{n!} \int\limits_{\Omega_n} \prod_{i=1}^n \left( \frac{d^3 \rb_i}
{\Lambda_c^3} \right) \exp \left(- \beta {\cal H}_n\right) \:, 
\label{eq:zeta_n} \\
\beta{\cal H}_n&=& -Z\lambda_{\rm B} \sum_{i=1}^n \frac{1}{|\rb_i|} + 
\lambda_{\rm B} \sum_{i<j} \frac{1}{|\rb_i-\rb_j|} \:,
\eea
where the integration hypervolume $\Omega_n
\equiv\{ a<|\rb_i|<R_n, \forall i=1,\cdots,n\}$
depends on the cutoff $R_n$. To fix the value of the cutoff,  
we follow an argument similar to the one used by Bjerrum in his study of 
dipolar formation in simple electrolytes \cite{Bj26}.
Suppose that we have a $(n-1)$-cluster and we want to condense 
one more counterion to form a $n$-cluster. Because of the spherical
symmetry, the $(n-1)$ bound counterions contribute only to the
renormalization of the polyion charge. 
The probability of finding the $n^{\rm th}$ counterion
at a distance $r$ in the interval $dr$ is 
\bea
P(r)\,dr &\propto&
 dr\, r^2 \exp\left[ -\beta q \phi(r)\right] \nonumber\\
&=& dr\, r^2  \exp\left[ \left( Z-n+1\right)\lambda_{\rm B}/r\right] \:,
\eea
where $\phi(r)$ is
the electrostatic potential generated by the $(n-1)$-cluster.
The probability distribution $P(r)$ has a minimum at
\be
r=R_n=(Z-n+1)\frac{\lambda_{\rm B}}{2} \:,
\ee
which, following Bjerrum \cite{Bj26}, we shall 
interpret as the distance of closest approach at which the $n^{\rm th}$
counterion will become bound to the $(n-1)$-cluster. Since 
$R_n/a>1$, for a given reduced temperature, $T^{\ast}
=a/\lambda_{\rm B}$, there is a minimum value of the valence,
$Z_{\rm min}=2T^{\ast}-1$, below which no counterions can condense onto  
a polyion, that is, the thermal entropic energy, $2k_{\rm B}T$, 
will overcome the gain 
in electrostatic potential energy, $(Z-n+1)q^2/(Da)$, 
preventing the confinement
from taking effect. 

With the cutoff defined, we shall now attempt to calculate the internal 
partition function of a $n$-cluster, Eq.~(\ref{eq:zeta_n}).
 That, in itself, is a formidable
task, since it requires evaluation of the many-body
integrals~(\ref{eq:zeta_n}). Instead of
performing the integrations explicitly, we shall use the
local-density-functional
theory  
to find the free energy of the condensed layer of $n$ counterions, 
$\beta F_{n}^{\rm con}\equiv - \ln \zeta_n$. 
Let us define the local density
of counterions in the condensed layer of a $n$-cluster as 
\be
\varrho_c(\rb)=\sum_{i=1}^n \delta (\rb-\rb_i)\: .
\ee
Within the local-density 
approximation (LDA), the Helmholtz free-energy functional $\beta 
{\cal F}_{n}^{\rm con}[\varrho_c(\rb)]$ 
corresponding to density $\varrho_c(\rb)$  is 
\end{multicols}
\ruleleft
\medskip

\bea
\beta {\cal F}_{n}^{\rm con}[\varrho_c(\rb)] &=&
\int_{V_n} d^3\rb\, \varrho_c(\rb) \left\{ \ln 
\left[\varrho_c(\rb) \Lambda_c^3 \right]-1 \right\} 
- Z \lambda_{\rm B} \int_{V_n} d^3 \rb \, \frac{\varrho_c(\rb)}{|\rb|} + 
\frac12 \lambda_{\rm B} \int_{V_n} d^3 \rb\, d^3 \rb'\, 
\frac{\varrho_c(\rb) \varrho_c(\rb')}{|\rb-\rb'|} \nonumber\\
&&
+ \int_{V_n} d^3 \rb\, \varrho_c (\rb) {\cal F}_{\rm corr}[\varrho_c(\rb)] \:,
\label{eq:cluster}
\eea
\vspace{-\baselineskip}

\ruleright
\begin{multicols}{2}
\noindent{}where the integrations are over the annulus $V_n \equiv
\{ a\le|\rb| \le R_n \}$ and ${\cal F}_{\rm corr}$ is given by 
Eq.~(\ref{eqn:ocp1}).
The first term in (\ref{eq:cluster}) corresponds to the usual
ideal-gas contribution, the second and the third terms are due to 
the electrostatic interactions and the last term is the result of the
correlations between the bounded counterions, for which we use the 
expression of the OCP theory \cite{No84}.
The equilibrium configuration, $\rho_c(\rb) = \left\langle
\varrho_c(\rb) \right\rangle$, is the one that minimizes the 
free-energy functional  $\beta {\cal F}_{n}^{\rm con}[\varrho_c(\rb)]$ under 
the constraint 
\be
\int_{V_n} d^3 \rb \, \rho_c(\rb) =n \:. 
\ee 
This minimization procedure leads to the Boltzmann distribution for the 
density profile,
\be
\rho_c( \rb)= 
\fracd{n  
\exp \left[-\mu_{\rm corr}(\rb) -
\beta q \psi(\rb) \right]}
{\int_{V_n} d^3 \rb'\, 
\exp \left[-\mu_{\rm corr}(\rb') -
\beta q  \psi(\rb') \right]} \label{eqn:boltzmann} \:,
\ee
where the electrostatic and 
the excess chemical potentials are given, respectively, by
\bea
\psi(\rb) &=& -\frac{Zq}{D|\rb|} + q \int_{V_n} d^3 \rb' \,
 \frac{\rho_c(\rb')}{D|\rb-\rb'|} \:, \\
\mu_{\rm corr} (\rb) &=&
{\cal F}_{\rm corr}[\rho_c(\rb)] +
\rho_c(\rb) \frac{\delta {\cal F}_{\rm corr}[\rho_c(\rb)] }
{\delta \rho_c(\rb)} \nonumber\\
&=&
 {\cal F}_{\rm corr}[\rho_c(\rb)] + \frac1{12} 
\left\{1- \omega^2 [\rho_c(\rb)]\right\} \:, 
\eea
with $\omega$ given by Eq.~(\ref{eqn:ocp2}).
On the other hand, the electrostatic potential and the 
total charge density satisfy the Poisson equation,  
\be
-\nabla^2 \psi(\rb) = \nabla \cdot \Eb (\rb)= \frac{4\pi}{D}
\left[ \sigma_0\, \delta(|\rb|-a)  + q \rho_c(\rb) \right]
\:,
\label{eqn:poisson}
\ee
where $\Eb (\rb)=-\nabla \psi(\rb)$ is the electric field at the 
point $\rb$.  Inserting  (\ref{eqn:boltzmann}) into (\ref{eqn:poisson})
we find a Poisson-Boltzmann-{\it like}\/ equation, which is a 
second-order nonlinear differential equation for the electrostatic
potential $\psi(\rb)$.
It should be remarked that, neglecting the correlation term, 
we regain the {\it usual}\/ Poisson-Boltzmann equation, 
$\nabla^2 \psi(\rb) \propto \exp\left[
-\beta q \psi(\rb) \right]$.

Since the boundary conditions are given in terms of the electric
field strength,
\be 
E(|\rb|=a)= - \fracd{Zq}{D a^2}\ \mbox{ and }\ 
E(|\rb|=R_n)=- \fracd{(Z-n)q}{D R_n^2} \:,
\ee
to perform the numerical calculations it is convenient to rewrite the 
equations in terms of this variable.
We now take advantage of the spherical symmetry of the system to eliminate
the angular dependence of the equations, that is, we  replace 
$\rb$ by $r=|\rb|$ in Eqs.~(\ref{eqn:boltzmann}) to (\ref{eqn:poisson}). 
The electric field also has only a spherically symmetric radial 
component, so that $\Eb(\rb)=E(r)\fracd{\rb}{r}$.    
Integrating the Poisson equation~(\ref{eqn:poisson}) over a sphere of 
radius $r$ and using the divergence  theorem, we obtain a relation 
for the electric field strength $E(r)$,
\end{multicols}
\be
\int\limits_{|\bms{r}'|<r}d^3\rb'\,\nabla\cdot \Eb(\rb')=
\int\limits_{|\bms{r}'|=r}d\bm{S}'\cdot \Eb(\rb')
=4\pi r^2 E(r)
= -\frac{4\pi q}{D} \left[Z - \int\limits_{|\bms{r}'|<r} d^3\rb'\,
\rho_c(\rb') \right] \:. \label{eqn:elec_field}
\ee

Inserting (\ref{eqn:boltzmann}) into (\ref{eqn:elec_field})
we obtain an integro-differential equation for the electric field,

\bea
E(r)&=&-\frac{q}{D r^2} \left\{ Z- n\:
\fracd{\int_{a}^{r} dr'\, r'^2 
\exp \left[ -\mu_{\rm corr}(r')
+ \beta q\int_{a}^{r'}dr''
\,{E}(r'')\right] }
{\int_{a}^{R_n} dr'\, r'^2
\exp \left[-\mu_{\rm corr}(r')
+\beta q  \int_{a}^{r'}dr''\,{E}(r'')\right]} \right\}
\:,
\eea

\ruleright
\begin{multicols}{2}
\noindent{}where we have chosen the gauge in which $\psi(r=a)=0$, and 
the density profile, $\rho_c(r)$, which is necessary to
evaluate $\mu_{\rm corr}(r)$, is also written in terms of 
the electric field,
\be
\rho_c(r) = \frac{D}{4\pi q} \nabla\cdot\Eb =
\frac{D}{4\pi qr^2} \frac{d}{dr} 
\left[r^2 E(r) \right]   \:. \label{eqn:density}
\ee 
The integro-differential 
equation was solved iteratively to obtain the electric field $E(r)$. 
The charge density $\rho_c(r)$ is then calculated
using Eq.~(\ref{eqn:density}).
Finally, the internal free energy of a $n$-cluster can be expressed in
terms of the charge density and the electric field to be 
\end{multicols}
\ruleleft
\medskip

\bea
\beta F_n^{\rm con}= \beta {\cal F}_{n}^{\rm con}
[\rho_c(r)] &=& 
\int_{V_n} d^3\rb\, \rho_c(r) \left\{ \ln 
\left[\rho_c(r) \Lambda_c^3 \right]-1 \right\}
+\frac{\beta D}{8\pi} \int_{V_n} d^3\rb\, E^2(r)  
+ \frac{(Z-n)^2 \lambda_{\rm B}} {2 R_n} - \frac{Z^2 \lambda_{\rm B}}{2 a}
\nonumber\\
&&
+ \int_{V_n} d^3 \rb\, \rho_c (r) {\cal F}_{\rm corr}[\rho_c(r)]
 \:, 
\eea

\ruleright
\begin{multicols}{2}
\noindent{}while the internal partition function of a $n$-cluster is  
$\zeta_n=\exp\left({-\beta F_{n}^{\rm con}}\right)$.
  
\section{Thermodynamic properties}

The total Helmholtz free energy of the polyelectrolyte solution is a sum of the 
entropic and the electrostatic contributions,
\be
f (\rho_f,\{ \rho_n \})= 
{f}^{\rm mix} + {f}^{\rm PC} + {f}^{\rm PP} + {f}^{\rm CC} \:.
\ee
Minimization of the total Helmholtz free energy under the constraints 
of fixed number of polyions and counterions leads to the law of mass 
action,
\be
\mu_0+n\mu_f=\mu_n \:, 
\ee
where the chemical potential of a specie $s$ is
 $\mu_s=-\partial f/\partial \rho_s$.
This results in a set of $Z$ coupled nonlinear algebraic equations for
the densities $\rho_n$, whose form is suitable to the use of an 
iterative method. 
Starting from a uniform distribution of clusters $\{\rho_n\}$, 
we were able to solve
the coupled system numerically.
A sample of the  distributions obtained is presented 
in Fig.~$1$. Two features are worth remarking. 
The counterion condensation 
\begin{figure}[hb]
\begin{center}
\leavevmode
\epsfxsize=0.45\textwidth
\epsfbox[5 20 520 410]{"
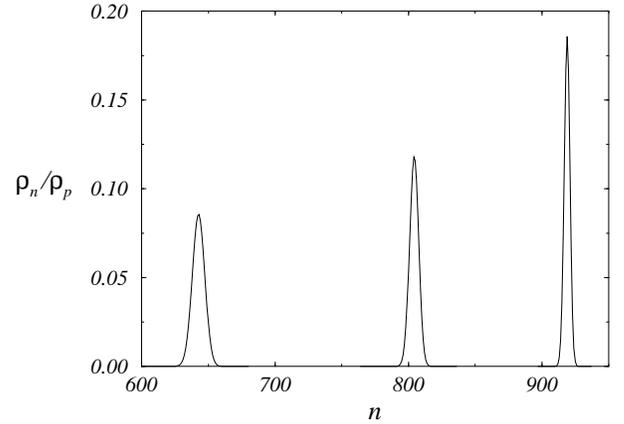"}
\end{center}
\begin{minipage}{0.48\textwidth}
\caption{Cluster-density distribution $\{\rho_n\}$ for $Z=1000$, 
volume fraction  $\phi=\frac43\pi a^3 N_p/V=0.01$, 
and various values of temperature. 
From left to the right, the values of the reduced temperature 
are $T^{\ast}=100,50,$ and $20$.}
\end{minipage}
\end{figure}
\noindent{}is more effective 
as the temperature decreases, and the width of the distribution 
is not very sensitive to the variations in temperature.
The pressure can be  obtained as a Legendre
transform of the Helmholtz free-energy density, 
\be
p=f(\rho_{f},\{\rho_n\})+\sum_{s} \mu_s\rho_s \:.
\ee

\begin{figure}[ht]
\begin{center}
\leavevmode
\epsfxsize=0.45\textwidth
\epsfbox[5 20 520 410]{"
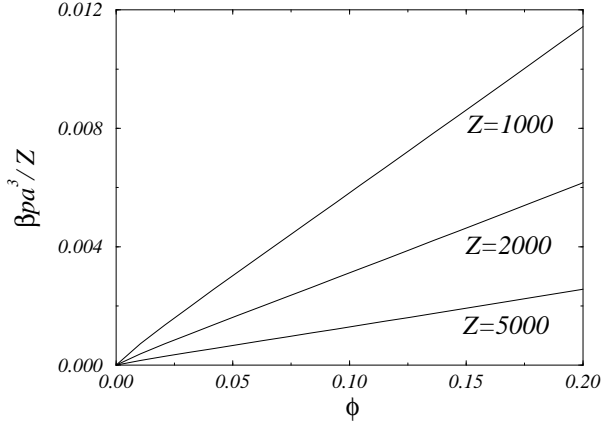"}
\end{center}
\begin{minipage}{0.48\textwidth}
\caption{Dependence of the dimensionless (total) pressure,
$\beta p a^3/Z$, on the volume fraction $\phi=\frac43\pi a^3 N_p/V$, 
for several values of $Z (1000,2000,5000)$ and $T^{\ast}=100$.}
\end{minipage}
\end{figure}

\begin{figure}[hb]
\begin{center}
\leavevmode
\epsfxsize=0.45\textwidth
\epsfbox[5 20 520 410]{"
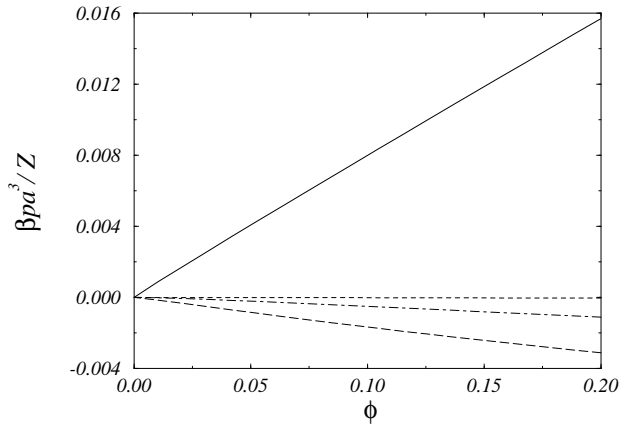"}
\end{center}
\begin{minipage}{0.48\textwidth}
\caption{Dependence of the various contributions to the
dimensionless (total) pressure, 
$\beta p a^3/Z$, on the volume fraction $\phi=\frac43\pi a^3 N_p/V$, 
for $Z=1000$, and $T^{\ast}=100$. Solid line: mixing (ideal-gas)
contribution; long-dashed line: polyion-counterion contribution; 
dot-dashed line: polyion-polyion contribution; dashed line:
counterion-counterion contribution.}
\end{minipage}
\end{figure}

\noindent{}In Fig.~2  we present the total pressure inside
the polyelectrolyte solution, which is a monotonically
increasing function of the density of polyions. 
In agreement with the current 
experimental and Monte Carlo results \cite{Vl89}, 
no evidence of any phase 
transition is encountered. 
To allow for a better appreciation of the relative importance of all the
terms, in Fig.~3 we present separately the various contributions to the 
total pressure. All the electrostatic terms give a {\it negative}\/ 
contribution to the total pressure, what can be interpreted as a form 
of an induced effective attraction between all the particles. 

\section*{Acknowledgments}

This work has been supported by the Brazilian agency CNPq (Conselho 
Nacional de Desenvolvimento Cient\'{\i}fico e Tecnol\'ogico).

\end{multicols}


\begin{thebibliography}{99}

\bibitem{Ma55} R. A. Marcus, J. Chem. Phys. {\bf 23}, 
1057 (1955);
S. Alexander, P. M. Chaikin, P. Grant, G. J.
Morales, P. Pincus, and D. Hone, 
J. Chem. Phys. {\bf  80}, 5776 (1984);
 R. D. Groot, J. Chem. Phys. {\bf 95}, 9191 (1991);
T. Gisler, S. F. Schulz, M. Borkovec, H. Sticher,
P. Schurtenberger, B. D'Aguanno, and R. Klein, J. Chem. Phys.  {\bf
101}, 9924
(1994);
V. Reus, L. Belloni, T. Zemb, N. Lutterbach, and H. Versmold, 
J. Phys. II France {\bf 7}, 603 (1997);
M. J. Stevens, M. L. Falk, and M. O. Robbins, 
J. Chem. Phys. {\bf 104}, 5209 (1996); 
L. B. Bhuiyan and C. W. Outhwaite, Mol. Phys. {\bf 87}, 625 (1996);
W. B. Russel and D. W. Benzing, J. Colloid
Interface Sci. {\bf 83}, 163 (1981); M. N. Tamashiro, 
Y. Levin, and M. C. Barbosa, to appear in J. Phys. II France.

\bibitem{Lo93} H. L\"owen, J.-P. Hansen, and P. A. Madden, 
J. Chem. Phys. {\bf 98}, 3275 (1993).

\bibitem{Fi93} M. E. Fisher and Y. Levin, Phys. Rev. Lett. {\bf
71},
3826 (1993); Y. Levin and M. E. Fisher, Physica  A {\bf 225}, 164
(1996); Y. Levin, X.-J. Li, and M. E. Fisher, Phys. Rev.
Lett. {\bf 73},
2716 (1994); 
 B. Guillot and Y. Guissani, Mol. Phys. {\bf 87}, 37
(1996). See also  G. R. Stell, K. C. Wu, and B. Larsen, Phys. Rev.
Lett. 
{\bf 37}, 1369 (1976); G. R. Stell, Phys. Rev. A {\bf 45}, 7628
(1992).

\bibitem{Bj26} N. Bjerrum, Kgl. Dan. Vidensk. Selsk. Mat.-Fys.
Medd. 
{\bf 7}, 1 (1926); H. Falkenhagen and W. Ebeling, in {\it Ionic
Interactions,} S. Petrucci (ed.)  
(Academic Press, New York, 1971), Vol. $1$: but 
note
slips in Eqs. $(49)$ and $(51)$.

\bibitem{DH23}(a) P. W. Debye and E. H\"uckel, Phys. Z. {\bf
24}, 185 (1923); (b) A good review is available in  D. A. McQuarrie,
{\it Statistical Mechanics} (Harper $\&$ Row, New York, 1976), 
Chap.~15.


\bibitem{Pa92}
A. Z. Panagiotopoulos, Fluid Phase Equil. {\bf 76}, 97 (1992);
{\bf 92}, 313 (1994);
  J. P. Valleau, J. Chem. Phys. {\bf 95}, 584 (1991).


\bibitem{LBT97}(a) Y. Levin, M. C. Barbosa, and M. N. Tamashiro,
Europhys. Lett. {\bf 41}, 123 (1998); (b)
Y. Levin, Europhys. Lett. {\bf 34}, 405  (1996); (c)
Y. Levin and M. C. Barbosa, J. Phys. II France {\bf 7}, 37 
(1997).

\bibitem{On33} L. Onsager, Chem. Rev. {\bf 13}, 73 (1933).     

\bibitem{DLVO} 
B. V. Derjaguin and L. Landau, Acta
Physicochimica
(USSR) {\bf 14}, 633 (1941);  E. J. W. Verwey and J. Th. G.
Overbeek, 
{\it Theory of the Stability
of Lyophobic Colloids} (Elsevier, Amsterdam, 1948);
M. Medina-Noyola and D. A. McQuarrie, J. Chem. Phys. {\bf 73},
6279 (1980);
X.-J. Li, Y. Levin, and M. E. Fisher, Europhys. Lett. {\bf 26},
683 (1994); 
M. E. Fisher, Y. Levin, and X.-J. Li, J. Chem. Phys. {\bf 101},
2273 (1994).

\bibitem{So84} We shall leave untouched the still controversial 
question of the existence of short-ranged attractive interactions 
between the two polyions [but see 
I. Sogami and N. Ise, J. Chem. Phys. {\bf 81}, 
6320 (1984), and Yan Levin (to be published)]. 
If this attraction exists, it will not affect 
strongly the thermodynamic properties of the solution, which are 
dominated by the counterions and their interactions with the polyions
(see Fig.~3).

\bibitem{No84}(a) S. Nordholm,  Chem. Phys. Lett. {\bf 105}, 302 (1984);
 (b) R. Penfold, S. Nordholm, B. J\"onsson, and C. E. Woodward, 
J. Chem. Phys.
{\bf 92}, 1915 (1990).

\bibitem{Vl89} 
V. Vlachy, C. H. Marshall, and A. D. J. Haymet, 
J. Am. Chem. Soc. {\bf 111}, 4160 (1989);
T. Palberg and M. W\"urth, Phys. Rev. Lett. {\bf 72},
786 (1994); J. C. Crocker and D. G. Grier, Phys. Rev. Lett. 
{\bf 73}, 352 (1994).   
\end{thebibliography}
\end{document}